\documentclass[a4paper,twoside,twocolumn,9pt]{article}
\usepackage[super,sort&compress,comma]{natbib} 
\usepackage[version=3]{mhchem}
\usepackage[left=1.5cm, right=1.5cm, top=1.785cm, bottom=2.0cm]{geometry}
\usepackage{balance}
\usepackage{times,mathptmx}
\usepackage{sectsty}
\usepackage{graphicx} 
\usepackage{lastpage}
\usepackage[format=plain,justification=justified,singlelinecheck=false,font={stretch=1.125,small,sf},labelfont=bf,labelsep=space]{caption}
\usepackage{float}
\usepackage{fancyhdr}
\usepackage{fnpos}
\usepackage[english]{babel}
\addto{\captionsenglish}{%
  
}
\usepackage{array}
\usepackage{droidsans}
\usepackage{charter}
\usepackage[T1]{fontenc}
\usepackage[usenames,dvipsnames]{xcolor}
\usepackage{setspace}
\usepackage[compact]{titlesec}
\usepackage{hyperref}

\newcommand{\be}{\begin{equation}}
\newcommand{\ee}{\end{equation}}
\newcommand{\ba}{\begin{array}}
\newcommand{\ea}{\end{array}}
\newcommand{\D}{\Delta}
\newcommand{\del}{\partial}
\definecolor{cream}{RGB}{222,217,201}

\begin{document}

\pagestyle{fancy}
\thispagestyle{plain}
\fancypagestyle{plain}{

\fancyhead[C]{\includegraphics[width=18.5cm]{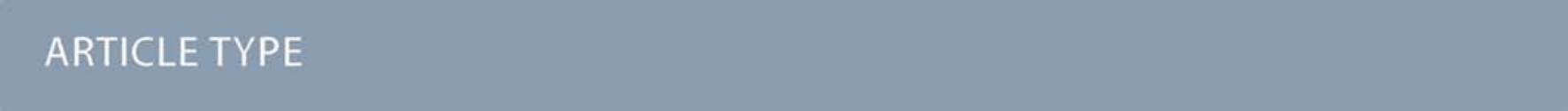}}
\fancyhead[L]{\hspace{0cm}\vspace{1.5cm}\includegraphics[height=30pt]{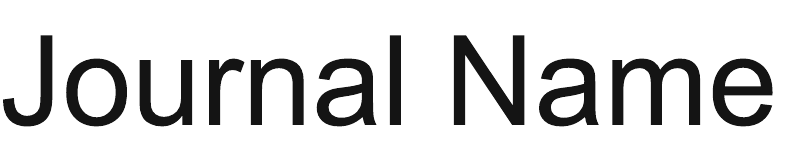}}
\fancyhead[R]{\hspace{0cm}\vspace{1.7cm}{www.rsc.org/XXXXXXX}}
\renewcommand{\headrulewidth}{0pt}
}

\makeFNbottom
\makeatletter
\renewcommand\LARGE{\@setfontsize\LARGE{15pt}{17}}
\renewcommand\Large{\@setfontsize\Large{12pt}{14}}
\renewcommand\large{\@setfontsize\large{10pt}{12}}
\renewcommand\footnotesize{\@setfontsize\footnotesize{7pt}{10}}
\makeatother

\renewcommand{\thefootnote}{\fnsymbol{footnote}}
\renewcommand\footnoterule{\vspace*{1pt}%
\color{cream}\hrule width 3.5in height 0.4pt \color{black}\vspace*{5pt}} 
\setcounter{secnumdepth}{5}

\makeatletter 
\renewcommand\@biblabel[1]{#1}            
\renewcommand\@makefntext[1]%
{\noindent\makebox[0pt][r]{\@thefnmark\,}#1}
\makeatother 
\renewcommand{\figurename}{\small{Fig.}~}
\sectionfont{\sffamily\Large}
\subsectionfont{\normalsize}
\subsubsectionfont{\bf}
\setstretch{1.125} 
\setlength{\skip\footins}{0.8cm}
\setlength{\footnotesep}{0.25cm}
\setlength{\jot}{10pt}
\titlespacing*{\section}{0pt}{4pt}{4pt}
\titlespacing*{\subsection}{0pt}{15pt}{1pt}

\fancyfoot{}
\fancyfoot[RO]{\footnotesize{\sffamily{1--\pageref{LastPage} ~\textbar  \hspace{2pt}\thepage}}}
\fancyfoot[LE]{\footnotesize{\sffamily{\thepage~\textbar\hspace{3.45cm} 1--\pageref{LastPage}}}}
\fancyhead{}
\renewcommand{\headrulewidth}{0pt} 
\renewcommand{\footrulewidth}{0pt}
\setlength{\arrayrulewidth}{1pt}
\setlength{\columnsep}{6.5mm}
\setlength\bibsep{1pt}

\makeatletter 
\newlength{\figrulesep} 
\setlength{\figrulesep}{0.5\textfloatsep} 

\newcommand{\topfigrule}{\vspace*{-1pt}%
\noindent{\color{cream}\rule[-\figrulesep]{\columnwidth}{1.5pt}} }

\newcommand{\botfigrule}{\vspace*{-2pt}%
\noindent{\color{cream}\rule[\figrulesep]{\columnwidth}{1.5pt}} }

\newcommand{\dblfigrule}{\vspace*{-1pt}%
\noindent{\color{cream}\rule[-\figrulesep]{\textwidth}{1.5pt}} }

\makeatother


\twocolumn[
  \begin{@twocolumnfalse}
\vspace{3cm}
\sffamily
\begin{tabular}{ p{18cm} }

\vspace{-1cm}

\noindent\LARGE{\textbf{Characterizing the fluid-matrix affinity in an organogel from the growth dynamics of oil stains on blotting paper}} \\
\vspace{0.3cm} \\

\noindent\large{Received Xth XXXXXXXXXX 20XX, Accepted Xth XXXXXXXXX 20XX} \\
\noindent\large{First published on the web Xth XXXXXXXXXX 20XX} \\
\vspace{0.3cm}

\noindent\large{Qierui Zhang,\textit{$^{a,b}$} Frieder Mugele,\textit{$^{a}$} Piet M. Lugt,\textit{$^{b,c}$} and Dirk van den Ende$^{\ast}$\textit{$^{a}$}} \\
\vspace{0.5cm}

\noindent\normalsize{Grease, as used for lubrication of rolling bearings, is a two-phase organogel that slowly releases oil from its gelator matrix. Because the rate of release determines the operation time of the bearing, we study this release process by measuring the amount of extracted oil as a function of time, while we use absorbing paper, to speed up the process. The oil concentration in the resulting stain is determined by measuring the attenuation of light transmitted through the paper, using a modified Lambert-Beer law. For grease the timescale for paper imbibition is typically 2 orders of magnitude larger than for a bare drop of the same base oil. This difference results from the high affinity, {\it i.e.} wetting energy per unit volume, of the oil for the grease matrix. To quantify this affinity, we developed a Washburn-like model describing the oil flow from the porous grease into the paper pores. The stain radius versus time curves for greases at various levels of oil content collapse onto a single master curve, which allows us to extract a characteristic spreading time and the corresponding oil-matrix affinity. Lowering the oil content results in a small increase of the oil-matrix affinity yet in a significant change in the spreading timescale. Even an affinity increase by a few per mill doubles the timescale.} \\

\end{tabular}

 \end{@twocolumnfalse} \vspace{0.6cm}

  ]

\renewcommand*\rmdefault{bch}\normalfont\upshape
\rmfamily
\section*{}
\vspace{-1cm}


\footnotetext{\textit{$^{a}$~Physics of Complex Fluids group, Faculty of Science and Technology, University of Twente, P.O. box 217, 7500AE, Enschede, The Netherlands. Email: h.t.m.vandenende@utwente.nl}}
\footnotetext{\textit{$^{b}$~Laboratory of Surface Technology and Tribology, Faculty of Engineering Technology, University of Twente, P.O. box 217, 7500AE, Enschede, The Netherlands.}}
\footnotetext{\textit{$^{c}$~SKF Research \& Technology Development, Postbus 2350, Nieuwegein, The Netherlands.}}



\section{Introduction}

Grease is a soft material that consists of an organogelator
network formed by self-assembled fibers or aggregates of platelets. The network serves as a matrix retaining oil up to a typical volume fraction of 80-90\% \cite{Lugt2009, Lugt2012}. Depending on the type of grease, the network is formed by fibers ranging from 0.1 to 0.5 $\mu m$ in diameter and 0.2-5 $\mu m$ in length, or by clusters of micrometer sized platelets, and takes up 10-30\% of the volume fraction \cite{Cyriac2016, Roman2016}. Grease is widely used in rolling bearings as a reservoir to retain and slowly release oil for the lubrication of the bearing. The oil should be released very gradually, ideally throughout the full life time of the bearing, {\it i.e.} typically several years. This process is very similar to the gradual release of liquid from food products \cite{Walstra1993}, responsive gels and (de-) swelling polymer brushes with applications {\it e.g.} in drug release, lubrication, anti-icing, self-cleaning surfaces \cite{Liu2013, Yao2014, Urata2015} and in art conservation \cite{Carretti2010, Pianorsi2017}. Despite this wide range of applications and the technological relevance for lubrication, the physical properties of the grease/gel that govern the release of the liquid are not well understood, nor how they should be quantified. Qualitatively, this release process, which is also known in the engineering literature as "bleeding" \cite{Lugt2012}, is driven by wetting and capillary forces of the bearing surface \cite{Bremond2011}. These driving forces are counteracted by viscous forces and by the affinity of the oil for the grease matrix. 
In practice, maintenance engineers deposit patches of grease onto blotting paper at elevated temperature and observe how the oil stains the paper, in order to assess the state of the grease qualitatively~\cite{Noordover2016}.
A review of liquid spreading on substrates and in porous materials can be found in \cite{Rosenholm2015}.
 
In this work, we develop a quantitative scheme to depict the transfer of oil from the grease into porous paper and to extract an affinity parameter that describes the retention strength of the grease matrix. To this end, we monitor optically the growth of an oil stain after depositing a fixed amount of grease onto the paper in the course of a few hours. Transmitted light intensity profiles reveal the distribution of oil in the pores of the paper and allow to define an effective radius, based on the absorbed mass instead of the commonly used sharp-front approximation \cite{Danino1994}. To describe the time evolution of this radius we develop a Washburn-like model for the capillary flow from the grease matrix into the porous paper \cite{Washburn1921, Mendez2009}. The collapse of the spreading radius versus time curves onto a single master curve for greases with various levels of oil content allows us to extract a characteristic spreading time and an affinity parameter that characterizes the oil-organogel matrix interaction.

\section{Experimental approach}

\begin{figure}[htb]
\centering
\includegraphics[width=80mm]{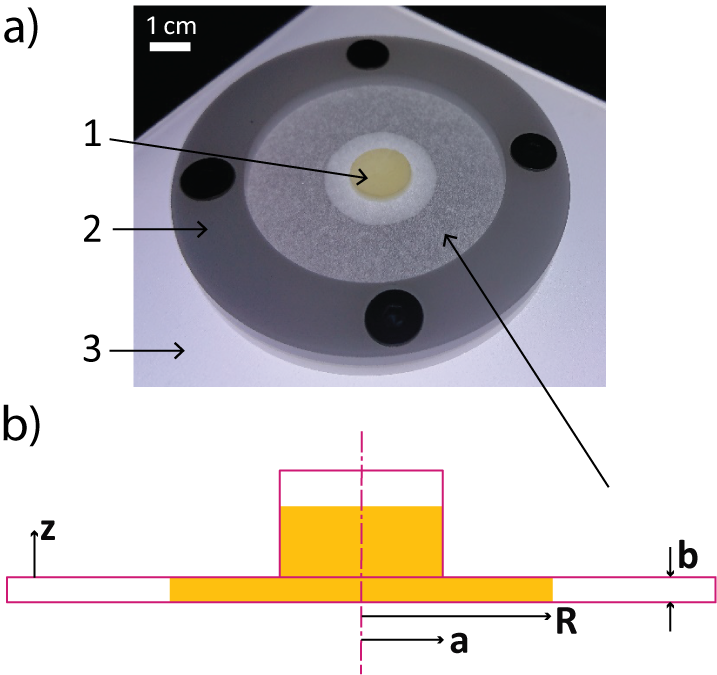}
\caption{\em a) Experimental setup and b) its schematic representation (not to scale) of the oil release test. 1: grease patch with radius $a$, 2: paper holder, 3: transmission illumination with a light diffuser, 4: paper with thickness $b$. The bright ring around the grease patch is the growing oil stain with radius $R$.}
\label{fig:setup}
\end{figure}

The experimental setup includes four basic elements: a free-standing sheet of blotting paper, on which the grease or the oil can be deposited, clamped between two plastic rings with an inner diameter of 20 mm; a LED pad (Metaphase) with a diffuser mounted below the sample holder, and a CCD camera (IDS, UI-3130-CP) to monitor the spreading process by recording the light transmitted perpendicularly through the sample with a frame rate of 0.6 (when extracting oil from oil drops) or 0.012 frames per second (from grease). Two types of paper were used, SKF MaPro kit test paper (denoted in the following as SKF paper) with a thickness of $182\ \rm \mu m$ and Whatman Chr1 chromatography paper (denoted as WCP paper) with a thickness of $167\ \rm \mu m$. Mercury porosimetry revealed that the SKF paper has a porosity of $(34.5 \pm 0.1)\ \%$ with pore sizes ranging from $0.2-15\ \rm \mu m$ in diameter. The corresponding values of the WCP paper are $(39.8\pm 0.2)\ \%$ and $1-30\ \rm \mu m$. 
To analyze the extraction of the oil from grease into the paper, cylindrical patches of grease with a diameter of 10 mm and a thickness of approximately 1.1 mm are deposited in the center of the paper using a glass mold and a spatula. Fig.~\ref{fig:setup} shows a side view image of a typical sample from an angle, with the grease patch in the center and a bright halo of enhanced transmission surrounding it due to oil that has been absorbed by the paper approximately 1-2 h after depositing the grease on the paper. The grease used in the present study consists of polyurea as a gelator with a volume fraction of 24\% and a synthetic ester oil with a density of $0.91\ \rm g/cm^3$ and a viscosity of $0.14\ \rm Pa\,s$ at room temperature. To investigate the effect of grease aging, measurements of oil extraction and spreading were performed both with neat grease as supplied as well as batches of grease from which the oil was partially removed by centrifugal filtration at $1500\ \rm rpm$ (432 g) at $20\ \rm ^\circ C$ using Whatman grade 1 filter paper. Depending on the duration of centrifugation, the grease was depleted to 76\% and 66\% of the original oil content, as determined by measuring the weight reduction of the centrifuged grease. Drops of the oil extracted by centrifugation with volumes between 3 and 18 $\rm \mu L$ were also deposited on blotting paper. Their spreading behavior was monitored as a reference and for calibration purposes to extract information about the permeability and the affinity between oil and paper. Top view images of the transmitted light (see Fig.~\ref{fig:snapshots}) were analyzed using standard image processing techniques. Specifically, the edge of the spreading oil stain was determined, after some background subtraction procedure, using a numerical routine for boundary detection in digital images. Subsequently, a least-squares circle fit to the numerically extracted edge was used to calculate the average front radius $R_{fr}$ of the oil stain.

\section{Results}
\subsection{Experimental phenomenology and physical approach} \label{ss:3.1}
\begin{figure*}[htb]
\centering
\includegraphics[width=150mm]{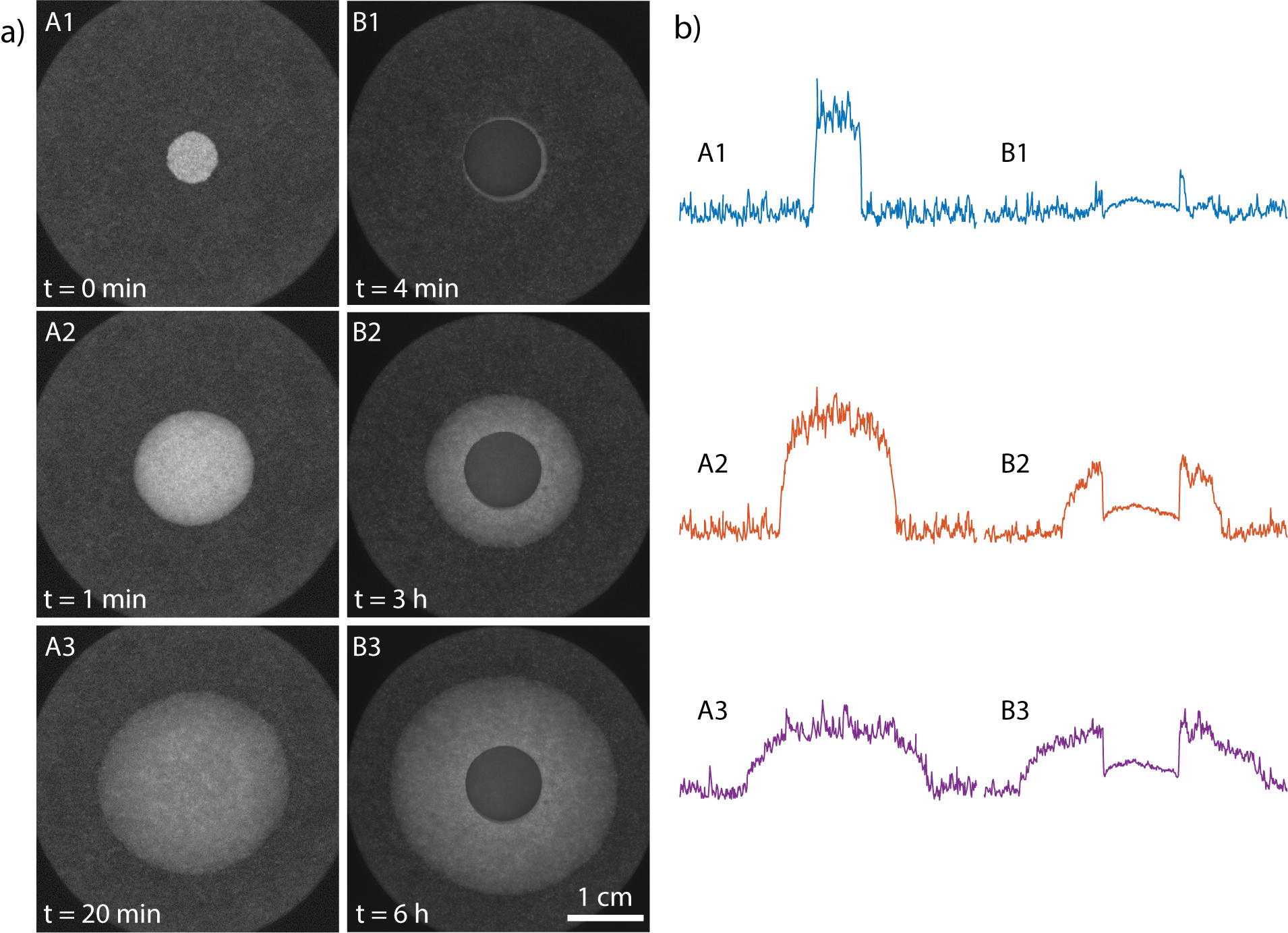}
\vspace{-3mm}
\caption{\em a) Time series of top-view snapshots of the oil stain due to an $10\ \rm\mu L$ oil drop (A1--A3) and a grease patch containing about 65 $\rm\mu L$ oil (B1--B3). b) The corresponding intensity profiles along a centerline through the stain. Note the dramatic reduction of spreading speed caused by the grease.}
\label{fig:snapshots}
\end{figure*}

\begin{figure}[h]
\centering
\includegraphics[width=80mm]{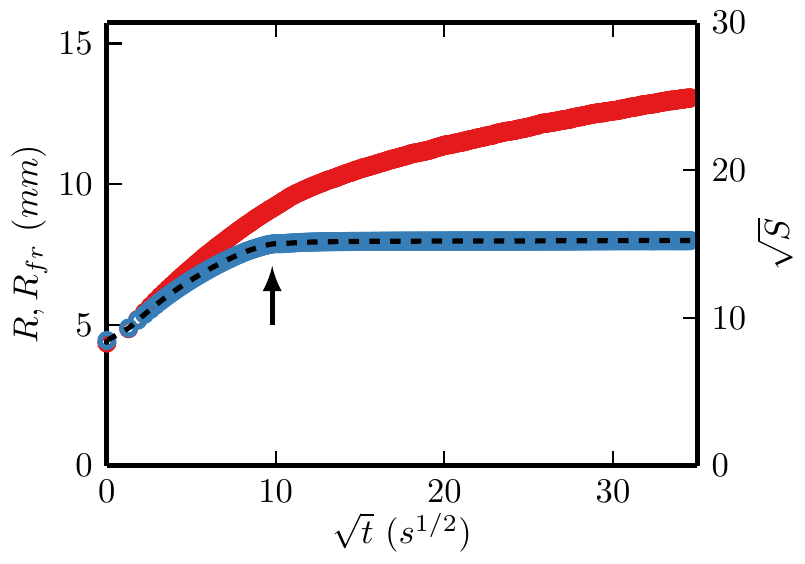}
\vspace{-3mm}
\caption{\em Front radius $R_{fr}$ as determined by tracing the edge of the stain (red circles) and the effective radius $R$ (blue circles) as determined by the integrated light intensity, both as a function of the square-root of time. On the right axis $S^{1/2}\ (\propto R)$ is given. The integrated intensity contrast $S$ is proportional to the amount of oil absorbed by the paper. The arrow indicates the point where the total drop is absorbed.}
\label{fig:SRf}
\end{figure}

Fig.~\ref{fig:snapshots} shows a series of characteristic snapshots of spreading oil stains from an oil drop (left column) and from a grease patch (right column) as well as series of corresponding intensity profiles. The oil drop was extracted from the same type of grease by centrifugal filtration. Plotting $R_{fr}$ as a function of time, see Fig.~\ref{fig:SRf}, reveals that the oil stain initially grows quickly and subsequently slows down, roughly in agreement with Washburn's law that predicts an advancement $\propto \sqrt{t}$. Note, however, that it takes approximately 10 times longer to reach a spreading radius of approximately $15\ \rm mm$ in the bottom row for the oil that is extracted from the grease patch as compared to the sessile drop. Since the oil and the paper are the same in both experiments, this retardation is direct evidence of the retention forces of the grease matrix that we aim to quantify in this study. 
The video images as well as the representative intensity cross sections shown in Fig.~\ref{fig:snapshots}b reveal that the spreading of the oil in the paper is in fact a two stage process. During the initial stages of oil spreading, the transmitted intensity in the center of the oil patch is essentially constant and drops sharply at $R_{fr}$ to the level of the surrounding dry paper. At later stages when the spreading rate has slowed down, however, the transition of the intensity at the edge of the stain becomes gradually more smeared out. Moreover, the total transmitted intensity in the center of the oil stain decreases. The occurrence of these two fluid spreading regimes in porous media has been discussed before by  Danino and Marmur \cite{Danino1994}  and has been attributed to the polydispersity of the pore sizes within the paper. During the first fast spreading regime of the oil, all pores become fully saturated with oil and the oil front is sharp. At the same time, the transmission of light is maximum because the oil greatly reduces the refractive index mismatch between the fibers of the paper and the pore space and thereby the scattering of light. As the spreading process slows down with time, however, the stronger suction power of the smaller pores overwhelms the lower hydraulic resistance of the larger pores. As a consequence, small pores ahead of the main part of the oil stain get filled and the front becomes increasingly dispersed. Danino and Marmur denoted this stage as redistribution regime to emphasize that oil is being redistributed from larger pores to smaller ones \cite{Danino1994}. 
For a finite reservoir such as our oil drops, this redistribution process implies that at very late stages, the fluid eventually fills the smallest pores across the entire sample. As a consequence, many of the initially filled larger pores in the center of the paper are gradually emptied again. In the experimental images this process also manifests itself in the decreasing transmitted intensity in the center of the oil stain due to the now again increasing light scattering from the larger pores. This trend is consistent with Gillespie's observations in his study of oil stains on paper \cite{Gillespie1958}. 

Oil stains spreading from a grease patch always spread much slower than the ones from the oil drop. As a consequence, the redistribution regime essentially sets in immediately as the oil stain starts to spread. Therefore, the front of the oil stain is rather gradual from the beginning and the transmitted intensity never reaches the maximum level of the fully oil-saturated paper (Fig.~\ref{fig:snapshots}b). 
While this redistribution process is very interesting as such and certainly deserves further investigation, it is also very complex and it is not the purpose of our present study, in which we aim to quantify the retention of the oil by the grease matrix. In that respect, the redistribution is a secondary process that takes place purely within the paper while our goal is to quantify the transfer of oil from the grease to the paper. Therefore, we will develop in the following a model that is essentially a Washburn-like imbibition model treating the grease and the paper as two separate porous media, each considered as homogeneous upon averaging over a sufficient number of pores. Each of these media is characterized by a certain affinity between the oil and the porous matrix, which is a characteristic energy per unit volume of the oil-filled medium. We expect that this energy will be dominated by the gain in interfacial energy upon wetting the porous matrix. Hence, it should depend on the interfacial tensions and the specific wetting area of the matrix. As a certain volume of oil is transferred from the grease into the paper, the system gains energy because the affinity of the paper is higher than the one of the grease matrix, as apparent from the spontaneous occurrence of the process in the experiments. This gain in wetting energy is opposed by viscous dissipation that we will quantify in a Darcy-approach, in principle in both media. Since the gain in wetting energy and the viscous friction within the paper will be identical for the spreading experiments from the oil drop and from the grease patch, we will strive to extract the grease properties from the retardation of the spreading as shown in Fig.~\ref{fig:snapshots}. 
With these ideas in mind, we assume that we can ignore the details of the redistribution of oil within the paper to a first approximation. To extract the quantities of interest, however, we do need to quantify the amount of oil in the paper based on our experimental data. While the front radius $R_{fr}$ can be extracted easily and robustly from all experimental data, as described above, it is clear from the lower-than-fully-oil-saturated transmitted intensity in the oil stains surrounding the grease patch that we cannot assume the pore space to be completely filled with oil all the way up to $R_{fr}$. Instead, we first need to develop a procedure to identify an effective radius $R$ up to which the paper can be assumed to be completely saturated and then proceed with a standard Washburn-Darcy-type model to describe the spreading dynamics and eventually extract the affinity of the oil to the grease. 

\subsection{Extracting the effective stain radius}\label{ss:2.3}
\subsubsection{Converting light intensity into oil concentration.~~}
\begin{figure}[h]
\centering
\includegraphics[width=80mm]{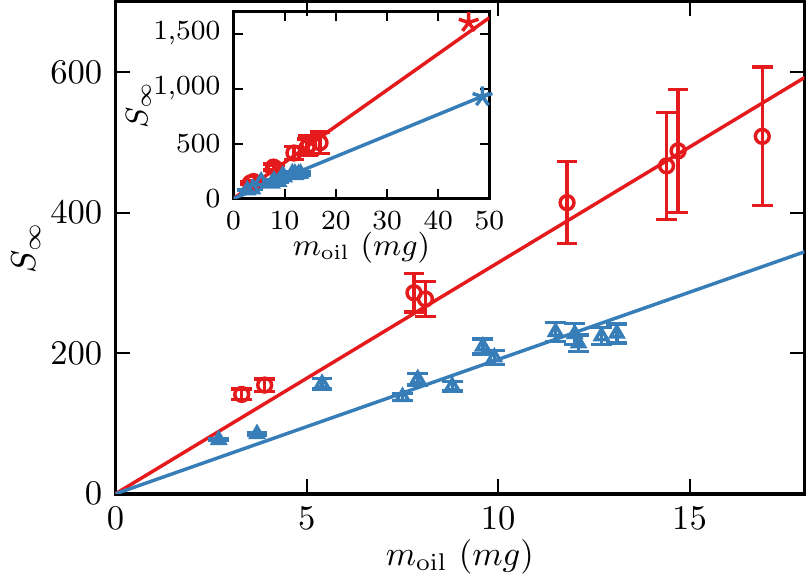}
\vspace{-3mm}
\caption{\em The integrated intensity contrast $S_\infty$ versus the mass of the absorbed oil drop, as  measured on SKF (red circles) and WCP paper (blue triangles). Solid lines: the linear fits. The star symbols in the inset show the saturation value $S_\infty$ when the considered area is fully saturated, versus the mass of oil absorbed in that area. The slopes indicate the intensity-mass conversion factor $m_\star^{-1}$: $33\pm2\ mg^{-1}$ and $19\pm1\ mg^{-1}$ for SKF and WCP paper, respectively.}
\label{fig:int2m}
\end{figure}

To convert the transmitted light intensity to the mass of oil in the paper, we implemented a modified Lambert-Beer relation. When light passes through the paper, the light intensity is reduced due to the volume fraction of the paper fibers, the empty pores in the paper and pores filled with oil. Thus, the intensity reduction $-dI$ when light passes through a layer of thickness $dz$ is described as:
\be
\frac{dI}{dz}=-qI
\label{eq:LBL}
\ee
where $q$ is the extinction coefficient. Assuming that the absorption and/or scattering of transmitted light scales not only with the density of pores as in the original Lambert-Beer Law \cite{vdHulst1981}, but also with a certain power $\beta$ of the radius $s$ of these pores, we define the  extinction coefficient $q$ in an ad-hoc manner as:
\be
q = \lambda_1 \int \alpha(s)n(s)s^\beta ds + \lambda_2 \int
[1-\alpha(s)]n(s)s^\beta ds + q_0
\ee
The three terms on the right hand side represent the attenuation contributions from the filled pores, the open pores and the paper fibers, respectively. Here $\alpha (s)$ is the fraction of pores with radius $s$ filled with oil and $1-\alpha (s)$ the fraction of open pores with radius $s$, while $n(s)ds$ is the number density of pores with a radius between $s$ and $s+ds$. The coefficients $\lambda_1$, $\lambda_2$ and $q_0$ are positive constants.
Because dry paper absorbs more light in transmission (looks darker) than wet paper, $\lambda_2$ should be larger than $\lambda_1$. Integrating Eq.~\ref{eq:LBL} over the thickness of the paper and normalizing the intensity with respect to the average background intensity of the dry paper, one obtains:
\be
\ln[I(\vec{r})/I_{bg}] = (\lambda_2-\lambda_1)b n_t\ \langle
s^\beta\alpha(\vec{r})\rangle
\label{eq:tildeI} 
\ee
where $I_{bg}$ is the transmitted intensity through dry paper, $n_t=\int n(s)ds$ the total number density of pores, $\langle s^\beta \alpha \rangle = n_t^{-1}\int s^\beta \alpha (\vec{r},s)\ n(s)ds$ the density weighted average of $s^\beta \alpha (\vec{r},s)$ and $b$ is the thickness of the paper. On the other hand, the volume fraction of oil is given by $\phi = \tfrac{4}{3}\pi n_t\langle s^3\alpha(\vec{r})\rangle$. For simplicity, we assume that: a) the pore size distribution is uniform, {\it i.e.} $n(s)=n_0$ for $s_{\min}<s<s_{\max}$, b) the radius $s_{\min}$ of the smallest pores is close to zero and c) the oil only fills and saturates all pores with a radius smaller than a critical value $s_\phi$ that depends on the local volume fraction of oil. With these assumptions, $\langle s^\beta \alpha \rangle\propto s_\phi^{\beta+1}$, where $\beta+1$ should be positive, and $s_\phi\propto\phi^{1/4}$. Substituting last two expressions in Eq.~\ref{eq:tildeI} one obtains:
\be
\phi(\vec{r})=\phi_\star\{\ln[I(\vec{r})/I_{bg}]\}^{4/(\beta+1)}
\ee
where $\phi_\star$ is a positive dimensionless constant that scales with $[b(\lambda_2-\lambda_1)]^{-4/(\beta+1)}$. Summing over the amount of oil in all pixels leads to the total mass of absorbed oil:
\be
m = \sum_{i,j} \rho b A_{px}\phi(i,j)=m_\star S
\label{eq:I2m} 
\ee
with $m_\star = \rho bA_{px}\phi_\star$, where $A_{px}$ is the area of a pixel and:
\be
S= \sum_{i,j} \{\ln[I(i,j)/I_{bg}]\}^c 
\label{eq:I2m-S}
\ee
where $c=4/(\beta+1)$. The integrated intensity contrast $S$ should be constant in the redistribution regime, when the drop is totally absorbed by the paper. Therefore we can determine the value of $c$ by adjusting the slope of the $S(t)$ curve in this regime to zero, which results in $c=1.67$. This value correspond to $\beta=1.4$ for the power-law dependence of the extinction coefficient on the pore radius. The conversion factor $m_\star$ is obtained by plotting the average plateau value of the integrated intensity contrast in the redistribution regime, $S_\infty$, versus the mass of the oil drops, see Fig.~\ref{fig:int2m}. The slope of the best fitting linear correlation, is equal to $m_\star^{-1}$. For both types of paper the experiments show a linear dependence. This linear relation, and so our modification of the Lambert-Beer law, is further validated by extrapolating the fits towards full saturation. At the center of the oil drop the paper is fully saturated. Therefore $S_\infty$ is determined by the transmitted light intensity at the center. The obtained value for $S_\infty$ is correctly predicted by the mass of oil as calculated for saturation of all pores in the paper, $m_s=\rho b \phi_p A$, where $\rho$ is the density of oil, $\phi_p$ is the porosity of the paper. The conversion factor $m_\star^{-1}$ is found to be $33 \pm 2\ \rm mg^{-1}$ for SKF paper and $19 \pm 1\ \rm mg^{-1}$ for WCP paper.

\subsubsection{Stain growth.~~}
From the absorbed mass of oil, $m$, we calculate an effective radius, $R$, which would be the radius of the stain if all pores within the stain were saturated with oil, {\it i.e.} $m=\rho\phi_p\pi b R^2$. Also this radius has been plotted as a function of $\sqrt{t}$ in Fig.~\ref{fig:SRf} (blue symbols). Comparing the effective radius $R$ with the front radius $R_{fr}$,  it is clear that we cannot use $R_{fr}$ to estimate the amount of liquid in the paper, because it still increases after the total drop has been taken up by the paper. But the radius $R$ reaches a plateau value once the drop is fully absorbed.

\begin{figure}[h]
\centering
{\large a)}\includegraphics[width=75mm]{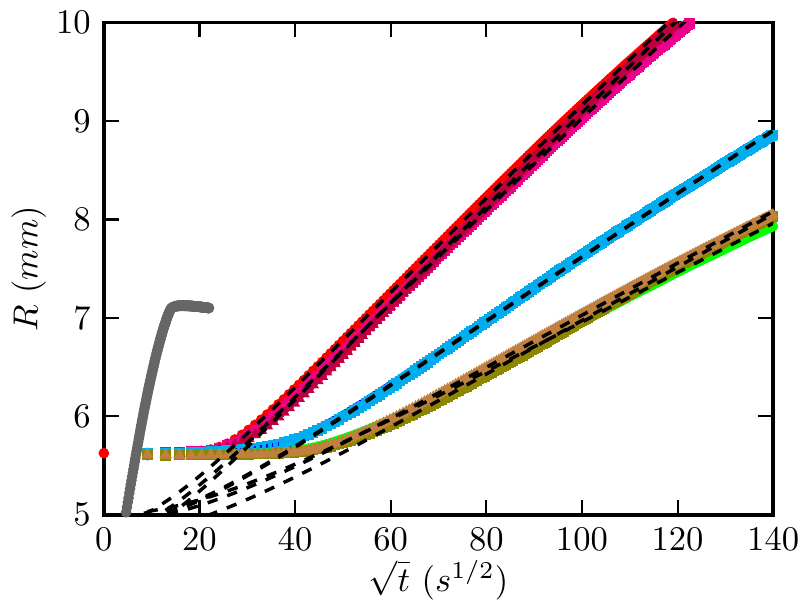}\quad 
{\large b)}\includegraphics[width=75mm]{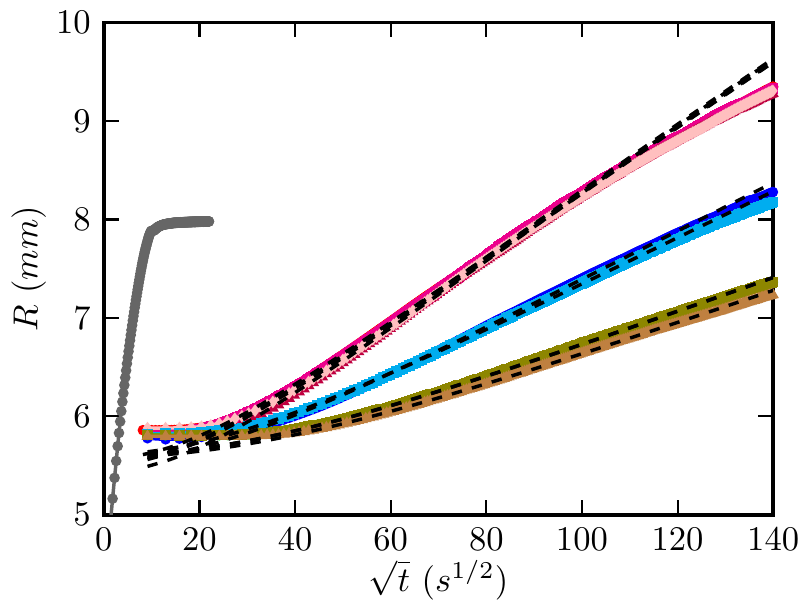}
\vspace{-3mm}
\caption{\em Effective radius $R$ versus square-root of time $\sqrt{t}$ for oil spreading in a) SKF paper and b) WCP paper, from a drop (leftmost gray curve) and from grease patches with 100\%, 76\% and 66\% (from upper left to lower right) of the initial oil content. The colored symbols represent the experimental data. For each concentration at least three different samples were used, all shown with their own color. The dashed lines represent the best fitting model curves.}
\label{fig:spreading}
\end{figure}

The oil stains obtained from grease patches are analyzed in the same manner, except for the circular region near the center, that is covered by the grease patch. Here, the patch itself causes an additional intensity reduction, as shown in Fig.~\ref{fig:snapshots}. In our analysis, this region is assumed to be saturated. This leads to a slight overestimation of the effective radius in the initial stage, because from the intensity profiles B1--B3 in Fig.~\ref{fig:snapshots} we observe that the intensity under the patch is not homogeneous, which implies that this area of the paper is not fully saturated. Therefore the total mass of absorbed oil is slightly overestimated and so the effective radius $R$. 
Despite this simplification, the model provides a quantitative comparison between the oil release dynamics of an oil drop and that of a grease patch. The effective radii of oil stains upon spreading from a drop or from a grease patch, for patches with various degrees of oil content, are plotted versus the square root of time in Fig.~\ref{fig:spreading}. Here $t=0$ is defined by the moment the grease patch is deposited on the blotting paper. For each concentration at least three different samples were used. All these curves are given in the figure, the overlap at each concentration indicates the reproducibility of the measurements.
Evidently, the timescale for spreading from an oil drop is much shorter than for spreading from a grease patch. Moreover, the more the grease patch is depleted, the slower the oil release. This observation suggests a gradually decreasing rate of oil release when a considerable amount of oil has already been extracted from the grease matrix as the grease ages in the bearing.

\section{Spreading model}
\subsection{Model description} \label{ss:modeling}
To analyze the oil release and spreading quantitatively, we developed a model that describes the oil flow from the grease patch into the blotting paper. It will predict the effective radius of the oil stain in the paper as a function of time.  

The rate of oil release is attributed mainly to the difference of the capillary pressures in the grease matrix and the paper which drives the release, and the permeability of paper and grease which causes a resistance to the oil flow. The capillary pressure can be obtained by balancing the surface wetting energy, $C_i\D \gamma_i\ \delta V $, in a certain volume $\delta V$ with the virtual work, $\phi_i\D p_i\ \delta V $,  done by the pressure drop over that volume:
\be
\D p_i = C^\star_i\D\gamma_i
\label{eq:Dp} 
\ee
where the subscript $i$ indicates the medium, grease matrix or paper. $\D p$ is the capillary pressure, being the pressure drop over the oil-air interface, and $C$ the specific wetting area, {\it i.e.} the surface area per unit volume of the porous medium. $\D \gamma$ is the difference in interface tension at the matrix-oil and the matrix-air interface and $\phi$ the porosity of the medium. Moreover, $C^\star=C/\phi$ is the surface area per unit volume of absorbed oil. Note that $\phi\delta V $ is the volume available to the oil. The product $C^\star\D\gamma$ is the wetting energy per unit volume of absorbed oil, discussed in section~\ref{ss:3.1}. We call $C^\star\D\gamma$ the wetting affinity. 

To model the dynamics of the oil release, we consider a slug of oil that penetrates from the grease matrix into the paper. As mentioned, both paper and grease matrix are modeled as a homogeneous and isotropic rigid porous medium. The geometry is schematically shown in Fig.~\ref{fig:setup}. Because the thickness of the paper is small compared to the base radius of the grease patch, we neglect the imbibition of oil in the vertical direction at the initial stage. In the grease matrix we assume a 1-D flow in the z-direction and in the paper a 2-D radial flow. Based on mass conservation, the radial outflow in the paper is given by $v_r = -rv_z/2b$ for $r<a$ and:
\be
v_r=\frac{-a^2v_z}{2br}
\label{eq:msconsrv}
\ee 
for $r\ge a$, where $v_r$ is the average velocity in the radial direction, while $v_z$ is the average vertical velocity at the matrix-paper contact area with radius $a$. Incorporating Darcy's equation, $\del p/\del r=(-\mu/k)\ v_r$, we find for the pressure distribution in the paper:
\be
p(r) = \left\{ 
\ba{ll}
p_0+\dfrac{\mu a^2 v_z}{4bk_p}\,(r/a)^2 & \mbox{(r $<$ a)}\\ \\
p_0+\dfrac{\mu a^2 v_z}{4bk_p}(1+\ln(r/a)^2) & \mbox{(r $\geq$
a)}
\ea \right. 
\ee
where $\mu$ is the viscosity of the oil, $k_p$ the permeability of the paper and $p_0$ the pressure at $r=z=0$. We identify the oil front in the paper with the radius $R$ as defined in section \ref{ss:2.3} and introduce the
variable $\xi=(R/a)^2=m/m_0$ where $m_0=\rho\phi_p \pi a^2b$. Note that $\xi$ can be considered to be the dimensionless radius squared of the stain but also the dimensionless mass of oil in the stain. For $\xi>1$ we get for the pressure:
\be
p_\infty-\D p_p =p_0+\frac{\mu a^2 v_z}{4bk_p}(1+\ln\xi)
\label{eq:p_pp} 
\ee
where $p_\infty$ is the ambient pressure. Similarly, the oil flow in the grease is described by:
\be
p_\infty-\D p_g =p_b- H \frac{\mu v_z}{k_g} 
\label{eq:p_gr} 
\ee
where $H$ is the height of oil in the grease matrix, $k_g$ the permeability of the matrix, $\D p_g$ the capillary pressure in the grease matrix. Moreover, $p_b=\int_0^a p(r)\,2\pi rdr/(\pi a^2)$ is the average pressure at the lower side of the reservoir at $z=0$. Combining Eqs.~\ref{eq:p_pp} and \ref{eq:p_gr} leads to:
\be
\D P = -\frac{\mu a^2 v_z}{8bk_p}(1+2\ln\xi)-\frac{\mu Hv_z}{k_g}
\label{eq:bld-diff} 
\ee
where we define $\D P =\D p_p-\D p_g$. When $a^2 \gg bH_0$, the last term on the right hand side, containing the permeability of the grease matrix $k_g$, can be neglected. The speed of the oil front is given by $\dot{R}=v_r(R)/\phi_p$. Therefore one obtains from
Eqs.~\ref{eq:msconsrv} and \ref{eq:bld-diff} the differential equation:
\be
\frac{\del\xi}{\del\tau}=\frac{1}{1+2\ln\xi}
\ee
where $\tau = t/t_s$. The characteristic time $t_s$ is defined as:
\be
t_s = \frac{\mu\phi_p a^2}{8 k_p\D P}
\label{eq:t_s} \ee
Solving this differential equation with the initial condition $\xi(\tau=0)=1$, we get for $1\leq\xi\leq\xi_{\max}$:
\be
\tau = 2\xi\ln\xi-\xi+1
\label{eq:bld} \ee
where $\xi_{\max}=m_{\rm tot}/m_0$ is determined by the amount of oil $m_{\rm tot}$ available in the drop or grease patch. Eq.~\ref{eq:bld} will be used to determine $t_s$ (and in case of spreading from oil drops, also the radius $a$). With these values we determine $k_p\D P=\tfrac{1}{8}\mu\phi_p\ a^2/t_s$. 

\subsection{Characterizing the paper}
\label{ss:3.4}
From the oil spreading experiments we can only determine the product $k_p\D P$ and not $k_p$ and $\D P$ independently. Therefore a separate capillary rise experiment has been performed. Stripes of paper are dipped vertically in a reservoir filled with the same oil as used in the spreading experiments. 
\begin{figure}[htb]
\centering
\includegraphics[width=80mm]{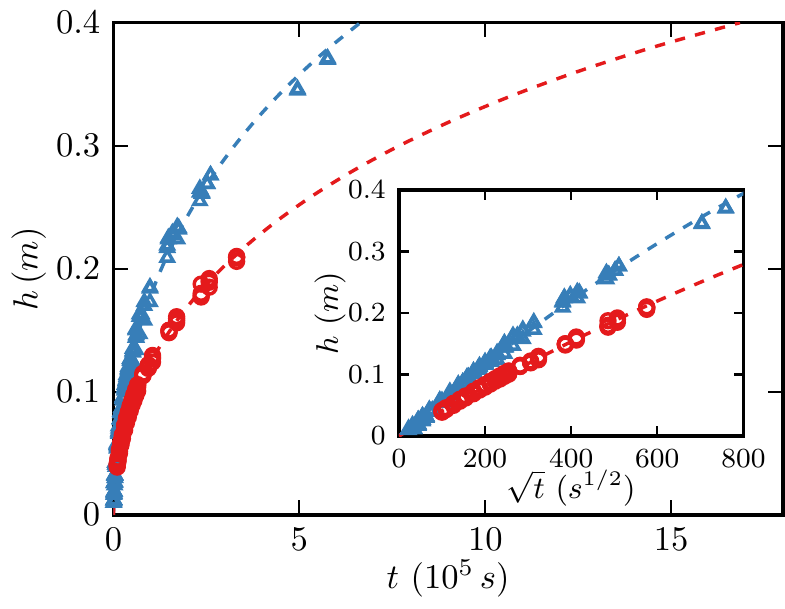} 
\vspace{-3mm}
\caption{\em Rising height of the oil front as a function of time for SKF (red symbols) and WCP paper (blue symbols). The insets show the rising height of the oil front versus square-root of time. The dashed lines represent the best fits of the capillary rise model.}
\label{fig:Caprise}
\end{figure}
In the capillary rise experiment, the width of the intensity gradient at the oil front is small, therefore only the height of the oil front is monitored over time. The final height $h_\infty$ of the oil front is determined by the capillary pressure $\D p_p$ in the paper, according to Jurin's law:
\be
\D p_p=\rho gh_\infty
\label{eq:Cap-hoo}\ee 
where $g$ is the acceleration due to gravity. When the oil front is still rising with velocity $\dot{h}=v_z/\phi_p$, the pressure difference $\D p_p$ is given by:
\be
\dfrac{\D p_p}{h} - \rho g = \dfrac{\mu \phi_p}{k_p}\dot{h}
\ee
Solving this equation, the time to reach a height $h$ is:
\be
\dfrac{t}{t_r}=-\ln\left(1-{h}/{h_\infty}\right)-{h}/{h_\infty}
\label{eq:Caprise} \ee
with the characteristic rising time:
\be
t_r = \frac{\mu\phi_ph_\infty}{\rho g k_p}
\label{eq:t_r} \ee
In the initial stage of capillary rise when $h \ll h_\infty$, Eq.~\ref{eq:Caprise} reduces to:
\be
h = w \sqrt{t}
\label{eq:Caprise-ini} \ee
where $w = h_\infty (2/t_r)^{1/2}$ is the initial slope of the rising height
versus the square-root of time. In principle we can extract from these measurements values for $\D p_p$ (from $h_\infty$) and $k_p$ (from $t_r$), separately. However, the deviation from the linear behavior is small, as we can observe from the inset of Fig.~\ref{fig:Caprise}, where the rising height $h$ has been plotted versus $\sqrt{t}$. Therefore, a reliable value for $h_\infty$is hard to obtain due to lack of data in the latest stages of the capillary rise. Measuring the rising height over longer time would improve the accuracy, but side effects such as evaporation and paper degradation due to swelling, limit the maximum observation time. In order to obtain more accurate results, we fix the value for the initial slope to:
\be
w = \sqrt{2k_p\D p_p/(\mu\phi_p)}
\label{eq:Caprise-w2} \ee
where the value for $k_p\D p_p/(\mu\phi_p)=\tfrac{1}{8}\,a^2/t_s$ has been taken from the oil spreading tests which will be discussed in the next section. Using Eq.~\ref{eq:Caprise-w2} as a constraint, we fit Eq.~\ref{eq:Caprise} to the capillary rise data using only the final height $h_\infty$ as fit parameter. The best fitting curves have been shown in Fig.~\ref{fig:Caprise}. For the final height we obtain $63\pm17\ \rm cm$ for SKF paper and $68\pm15\ \rm cm$ for WCP paper, corresponding to a capillary pressure of around $6$ kPa for both papers and a permeability of $0.81\pm0.26\ \rm \mu m^2$ and $1.82\pm0.63\ \rm\mu m^2$ for SKF and WCP paper, respectively. These data have been summarized in Table \ref{tbl:summary}.

\section{Discussion}
\begin{figure}[htb]
\centering
{\large a)}\includegraphics[width=75mm]{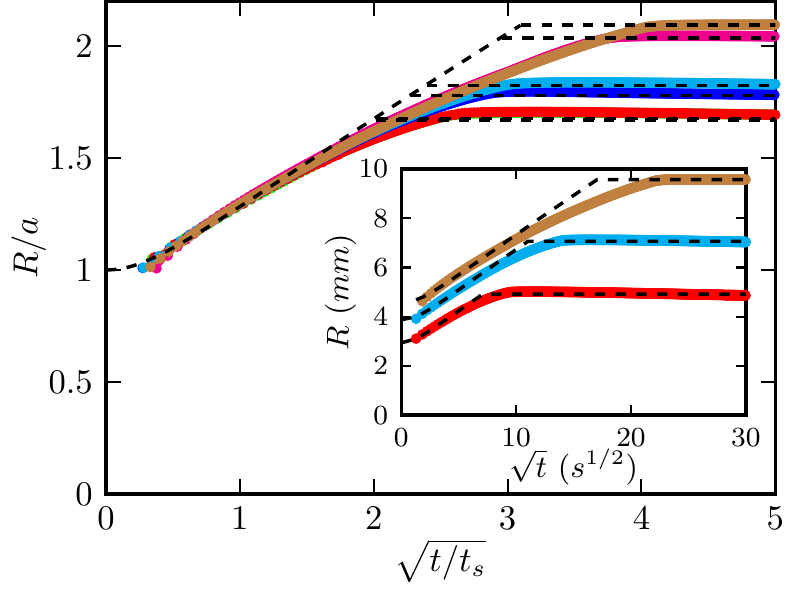} \quad 
{\large b)}\includegraphics[width=75mm]{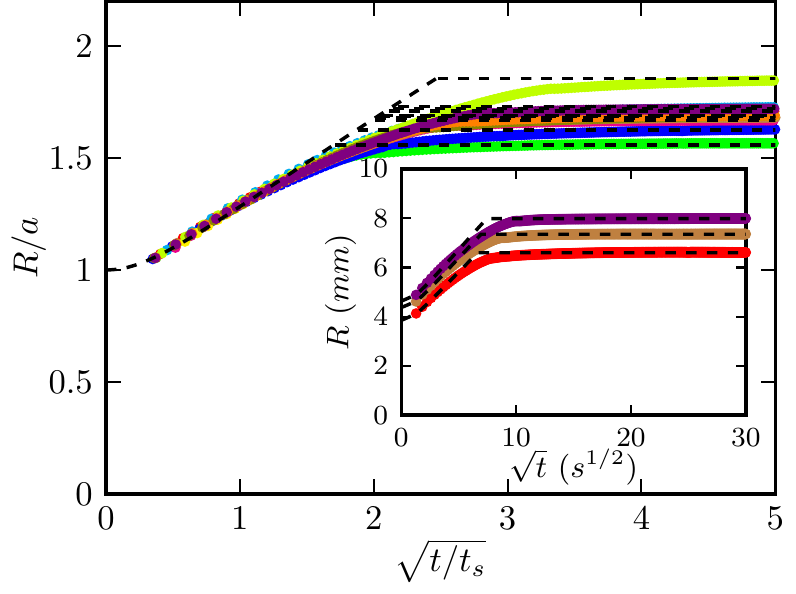}
\vspace{-3mm}
\caption{\em Normalized effective radius $R/a$ of the oil stain versus square-root of the normalized time $\sqrt{t/t_s}$ for the spreading from oil drops with various volumes on SKF paper (panel a) and WCP paper (panel b). Insets: Time dependence of the radius $R$ for oil drops of various volumes. Dashed lines: fits of the oil spreading model.}
\label{fig:oil}
\end{figure}

\subsubsection{Spreading from a drop.~~}
To validate our model calculations, Eq.~\ref{eq:bld} has been fitted to the oil spreading data. The obtained fits are shown in Fig.~\ref{fig:oil}. For short times, $t/t_s\le 1$, the experimental curves nicely collapse onto the expected master curve, described by Eq.~\ref{eq:bld}. From the fit we obtain the spreading time $t_s$ and the initial radius $a$ of the apparent three phase contact line of the drop on the blotting paper. 
With these values we determine $w=(\frac{1}{4}a^2/t_s)^{1/2}$, used for analyzing the capillary rise data described in section~\ref{ss:3.4}, and $k_p\D p_p$, see Table \ref{tbl:summary}. For long times, $t/t_s\gg1$, the experimental curves reach a plateau because the drop is fully absorbed by the paper and the influx of oil stops. The value of this plateau radius $R_\infty=a\,\xi_\infty^{1/2}$ matches correctly with the mass of the initial oil drop, indicated by the horizontal dotted line, see also section~\ref{ss:modeling}. The insets of the graphs show the corresponding radius $R$ versus $\sqrt{t}$ behavior for some of the individual curves, showing the agreement between the model calculation and the experimental data even more clearly.

\begin{table}[htb]
\small
  \caption{\ Values of the fit parameters as obtained from the capillary rise and oil spreading tests, together with  the paper and grease properties, as derived from them}
  \label{tbl:summary}
  \begin{tabular*}{0.48\textwidth}{@{\extracolsep{\fill}}lll}
  \hline
                                & {SKF}           & {WCP}                \\
  \hline
  Capillary rise                &                 &                      \\
  $h_\infty\,({\rm cm})$        & $63\pm17$       & $68\pm15$            \\
  $\D p_p\,({\rm kPa})$         & $5.6\pm1.5$     & $6.1\pm1.3$          \\
  $k_p\,({\rm \mu m^2})$        & $0.81\pm0.26$   & $1.82\pm0.63$        \\
  \\
  Base oil                      &                 &                      \\
  $a^2/t_s\,({\rm mm^{2}/s})$   & $0.70\pm0.03$   & $1.6\pm0.2$          \\
  $k_p \D p_p\,({\rm \mu m^2\ kPa})$  & $4.5\pm0.2$\ & $11.0\pm1.4$      \\
  \\
  Grease                        &                 &                      \\
                                &\textit{100\% oil cont.}    &           \\ 
  $a^2/t_s\,({\rm mm^{2}/s})$   & $(164\pm5)10^{-4}$ & $(73\pm2)10^{-4}$ \\
  $\D p_p-\D p_g\,({\rm Pa})$   & $129\pm42$      & $28\pm10$            \\
                                &\textit{76\% oil cont.}     &           \\
  $a^2/t_s\,({\rm mm^{2}/s})$   & $(75\pm1)10^{-4}$  & $(37\pm2)10^{-4}$ \\
  $\D p_p-\D p_g\,({\rm Pa})$   & $59\pm19$       & $14\pm5$             \\
                                &\textit{66\% oil cont.}     &           \\
  $a^2/t_s\,({\rm mm^{2}/s})$   & $(49\pm4)10^{-4}$ & $(19\pm1)10^{-4}$  \\
  $\D p_p-\D p_g\,({\rm Pa})$   & $38\pm13$       & $7\pm2$              \\
  \hline
  \end{tabular*}\\
\end{table}

\subsubsection{Oil-matrix affinity.~~}
\begin{figure}[htb]
\centering
\includegraphics[width=75mm]{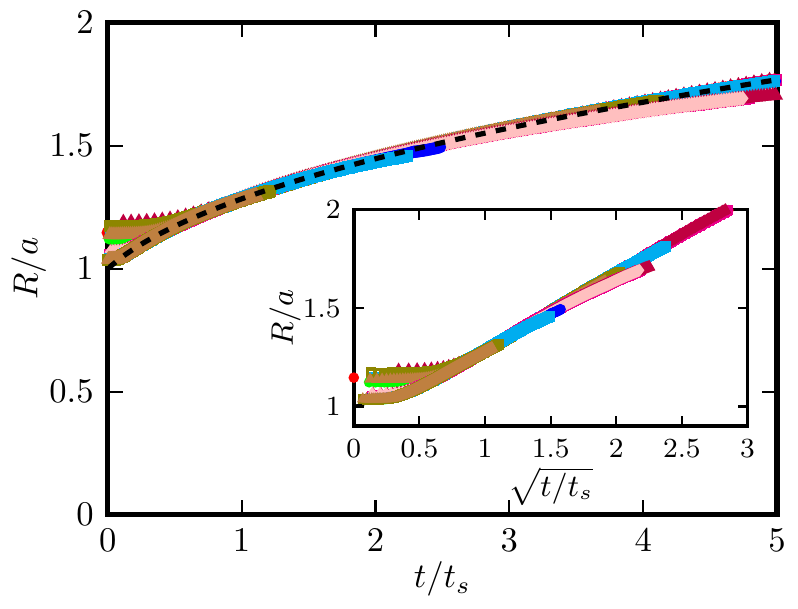}
\vspace{-3mm}
\caption{\em Normalized radius $R/a$ of the oil stain as a function of normalized time, $t/t_s$ for oil release from a grease patch, showing the collapse of all experimental data, for both SKF and WCP paper, to a single master curve. The inset shows $R/a$ versus $\sqrt{t/t_s}$. Color coding corresponds with Fig.~5.}
\label{fig:grease}
\end{figure}

In Fig.~\ref{fig:spreading} the spreading data from greases with varying oil content are shown as $R$ versus $\sqrt{t}$ for both SKF and WCP paper. 
When fitting the spreading model to these data, the initial stage for $t/t_s<0.5$ is not considered, because in this regime the amount of adsorbed oil and so the radius $R$ is over-estimated, as explained in section \ref{ss:2.3}. 
In the time interval $0.5<(t/t_s)^{1/2}<3$ we can approximate the model curve, Eq.~\ref{eq:bld}, with an error less than 1\%, by the linear relation: 
\be
R\simeq R_0+c_1\ a t_s^{-1/2}\ (\sqrt{t}-c_2\sqrt{t_s}\,)
\ee
where $R_0=1.5\,a$, $c_1=0.39$ and $c_2=1.55$. The observed linear dependence is characteristic for Washburn-like imbibition \cite{Washburn1921}. Considering Fig.~\ref{fig:spreading}b we observe that for WCP paper at later times, $t/t_s>10$, the slope of the measured curves is slightly smaller than that of the model curve. This is possibly caused by a change in the effective permeability of the blotting paper, due to the redistribution of oil from larger to smaller pores. This has not been taken into account in the modeling. 
The resulting scaling is shown in Fig.~\ref{fig:grease}. Again, the experimental curves collapse onto the expected master curve (except for $t/t_s<0.5$) when we plot the dimensionless radius $\sqrt{\xi}=R/a$ versus the dimensionless time $\tau=t/t_s$ or versus $\sqrt{\tau}$ as shown in the inset of the figure. 
Using the fitted values for the characteristic radius $a$ (which deviate in all cases less than 10\% from the expected value of 5 mm) and spreading time $t_s$ we calculate $a^2/t_s$. These values have been given in Table \ref{tbl:summary}. 
From the table we read that the ratio $a^2/t_s$ decreases by at least a factor 40 when the oil drop is replaced by a grease patch. According to Eq.~\ref{eq:t_s} this ratio is proportional to the permeability $k_p$ of the paper and the difference $\D P=\D p_p-\D p_g$ of capillary pressures in the paper and in the grease matrix.  When a drop of oil is placed on the paper, the capillary pressure due to the curvature of the drop is negligible compared to the capillary pressure in the grease matrix. This results in a much faster oil release and spreading from the oil drops.
As discussed in section~\ref{ss:3.1} the capillary pressure $\D p_g$ can be identified with the oil-matrix wetting affinity. From this perspective, the release of oil from the grease is caused by a slightly higher affinity of the oil for the adsorbing paper than for the grease matrix. This difference in wetting affinity, as listed in Table \ref{tbl:summary}, is much smaller than the absolute affinity of the oil for the paper. Therefore, the oil imbibes the paper at a much slower rate from a grease patch than from an oil drop, as shown in Fig.~\ref{fig:spreading}. 

\begin{figure}[htb]
\centering
\includegraphics[width=80mm]{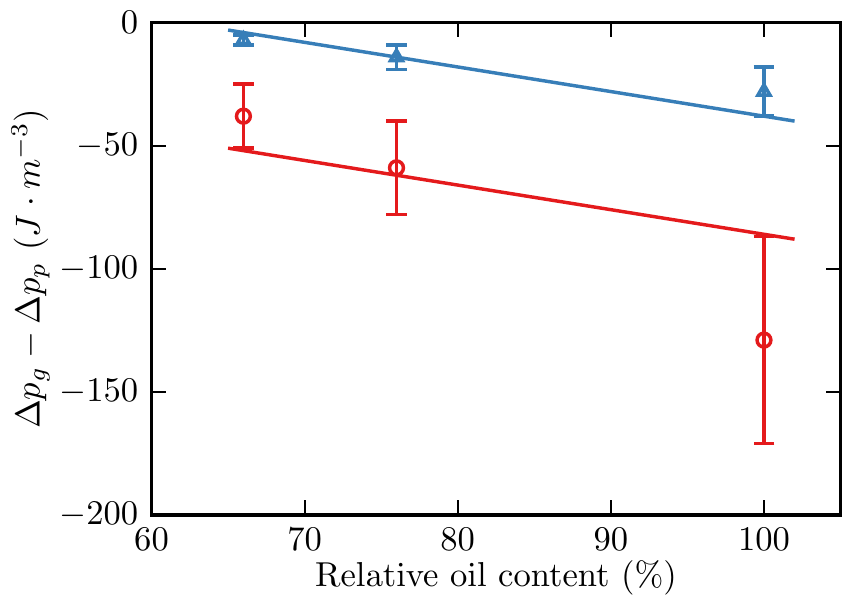}
\vspace{-3mm}
\caption{\em Difference between oil-grease and oil-paper wetting affinity, $\D p_g-\D p_p$, versus relative oil content of the grease patch for SKF (red symbols) and WCP paper (blue symbols).}
\label{fig:Dp}
\end{figure}

From the data in Table \ref{tbl:summary} we conclude that the wetting affinity of the oil for the grease matrix is: $\D p_g= 6\pm1\ \rm kJ/m^3$. For a fiber network with properties as mentioned in the introduction we estimate the specific wetting area of the grease as: 
\be
C^\star_g=\frac{1-\phi_g}{\phi_g}\ \frac{A_f}{V_f}=\frac{1-\phi_g}{\phi_g}\ \frac{2(a+l)}{al}
\ee
where $\phi_g=0.76$ is the pore volume fraction, $A_f$ the surface area of a single fiber, $V_f$ its volume. Assuming the fibers to be cylindrical with a characteristic radius $a$ and length $l$ of $0.2\ \rm \mu m$ and $3\ \rm \mu m$ \cite{Lugt2012}, we find $C^\star_g\simeq3\cdot 10^6\ \rm m^{-1}$. With Eq.~\ref{eq:Dp} we estimate $\D\gamma_g \simeq 2\ \rm mJ/m^2$. This is only a rough estimate but $\D\gamma_g$ has the right order of magnitude.    
For the blotting paper we also get the pore size distribution from the mercury porosimetry. Using these distributions we estimate for WCP paper $C^\star_g\simeq0.8\cdot 10^6\ \rm m^{-1}$ and for SKF paper $2.2\cdot 10^6\ \rm m^{-1}$. Therefore we obtain for $\D\gamma_p \simeq 7.6\ \rm mJ/m^2$ and $2.7\ \rm mJ/m^2$, respectively. Again this is in the range one should expect \cite{Meijer2000}.  
Unfortunately, the uncertainty in $\D p_g$ and  $\D p_p$ is quite large due to the large uncertainty in $h_\infty$ as obtained from the capillary rise experiment, see Table \ref{tbl:summary}, $\D h_\infty/h_\infty\simeq 0.25$. Considering the differences in $\D p_p-\D p_g$ for SKF and WCP paper, as presented in Fig.~\ref{fig:Dp}, we find: $\D p_p({\rm SKF})-\D p_p({\rm WCP})= 40\pm15\ \rm J/m^3$. It corresponds with the vertical distance between the trend lines in Fig.~\ref{fig:Dp}. Remarkably, the wetting affinities of both types of paper are equal within 1\% while their specific wetting areas and permeabilities differ by a factor two. 

\subsubsection{Aging of grease.~~}
As already explained, to mimic the aging of the grease in a rolling bearing, we intentionally depleted the oil from our grease samples before performing the spreading experiments. When the oil is partially removed from the grease, the spreading time increases considerably, as can be observed from Fig.~\ref{fig:spreading}, but the wetting affinity $\D p_g$ increases only slightly, as shown in Fig.~\ref{fig:Dp}, where $\D p_g-\D p_p$ as been plotted as a function of the relative oil content of the grease. 
Here we take the original fresh grease as a reference. When up to 34\% of the oil is removed from the matrix, the affinity increases, according to Fig.~\ref{fig:Dp}, by no more than $40\ {\rm J/m^3}$, which is approximately 1\% of the initial affinity. However, from  Fig.~\ref{fig:spreading} we observe that the spreading time on the absorbing paper increases by a factor~4. When a significant part of the oil is depleted, the grease matrix may partially collapse, leading to a denser matrix with a larger specific wetting area. The observed small increase in the oil-matrix affinity with decreasing oil content can be attributed to this change in the micro structure of the grease. 

\section{Conclusion}
To determine the wetting affinity of lubricating oil for its grease matrix, {\it i.e.} the total wetting energy per unit volume of grease, we extracted oil from the grease using blotting paper. We quantified this affinity by determining the capillary pressure in the grease matrix, compared with the capillary pressure in the paper and studied the spreading of the resulting oil stain in the paper.
We used an optical method to determine the local oil density in the stain and from that the total amount of oil extracted from the grease.
The amount of oil extracted by the paper is quantified by the attenuation of transmitted light intensity based on a modified Lambert-Beer law. With this method we measured the development of the stain radius as a function of time. 
The spreading time of the oil stain is modeled by calculating the effective stain radius $R$ as a function of time $t$, using a Washburn-like model, in which the oil flows from one porous medium (the grease matrix) into another (the absorbing paper). The characteristic time for the oil release and spreading is inversely proportional to the difference between the oil-paper and oil-matrix wetting affinity. 
Because the spreading time also depends on the permeability and porosity of the porous media involved, we determined in a separate experiment the permeability and affinity of the blotting paper, while its porosity was determined separately by mercury porosimetry.  
The model describes the observed spreading behavior very well, as can be concluded from the collapse of all experimental curves on a single master curve in Fig.~\ref{fig:grease}. 
The wetting affinity of the oil for the grease matrix of $6\ \rm kJ/m^3$ increases slightly as the oil content of the grease matrix is reduced. 
This gentle increase can be explained by a partially collapse of the matrix during oil depletion. Although it is small, less than 1\%, it leads to a significant increase of the spreading time of the oil stain, with approximately a factor 4. 

The obtained analytical and experimental insight in the oil-matrix wetting affinity and the resulting release dynamics of liquid from its matrix is useful not only for understanding the grease performance in rolling bearings, but also for applications that, for instance, utilize the syneresis of organogels \cite{Walstra1993}, or for constructing superhydrophobic, self-cleaning or anti-icing surfaces, where the formation of a thin liquid film on top of the gel coating is essential \cite{Liu2013, Yao2014, Urata2015}. It is also useful for optimizing cleaning procedures of delicate art work, or controlled drug delivery  \cite{Carretti2010, Pianorsi2017}.

\section*{Acknowledgements}
This research was carried out under project number S21.1.15582 in the framework of the Partnership Program of the Materials innovation institute M2i (www.m2i.nl) and the Technology Foundation STW (www.stw.nl), which is part of the Netherlands Organization for Scientific Research (www.nwo.nl). Part of the funding was provided by SKF Research and Technology Development. The authors wish to thank Dr. Andreas Gei\ss ler from the Macromolecular Chemistry and Paper Chemistry group, Technische Universit\"{a}t Darmstadt for performing the
mercury porosimetry measurements.



\balance

\bibliography{bob} 

\providecommand*{\mcitethebibliography}{\thebibliography}
\csname @ifundefined\endcsname{endmcitethebibliography}
{\let\endmcitethebibliography\endthebibliography}{}
\begin{mcitethebibliography}{19}
\providecommand*{\natexlab}[1]{#1}
\providecommand*{\mciteSetBstSublistMode}[1]{}
\providecommand*{\mciteSetBstMaxWidthForm}[2]{}
\providecommand*{\mciteBstWouldAddEndPuncttrue}
  {\def\EndOfBibitem{\unskip.}}
\providecommand*{\mciteBstWouldAddEndPunctfalse}
  {\let\EndOfBibitem\relax}
\providecommand*{\mciteSetBstMidEndSepPunct}[3]{}
\providecommand*{\mciteSetBstSublistLabelBeginEnd}[3]{}
\providecommand*{\EndOfBibitem}{}
\mciteSetBstSublistMode{f}
\mciteSetBstMaxWidthForm{subitem}
{(\emph{\alph{mcitesubitemcount}})}
\mciteSetBstSublistLabelBeginEnd{\mcitemaxwidthsubitemform\space}
{\relax}{\relax}

\bibitem[Lugt(2009)]{Lugt2009}
P.~M. Lugt, \emph{Tribology Transactions}, 2009, \textbf{52}, 470--480\relax
\mciteBstWouldAddEndPuncttrue
\mciteSetBstMidEndSepPunct{\mcitedefaultmidpunct}
{\mcitedefaultendpunct}{\mcitedefaultseppunct}\relax
\EndOfBibitem
\bibitem[Lugt(2012)]{Lugt2012}
P.~M. Lugt, \emph{Grease lubrication in rolling bearings}, John Wiley \& Sons,
  2012\relax
\mciteBstWouldAddEndPuncttrue
\mciteSetBstMidEndSepPunct{\mcitedefaultmidpunct}
{\mcitedefaultendpunct}{\mcitedefaultseppunct}\relax
\EndOfBibitem
\bibitem[Cyriac \emph{et~al.}(2016)Cyriac, Lugt, Bosman, Padberg, and
  Venner]{Cyriac2016}
F.~Cyriac, P.~M. Lugt, R.~Bosman, C.~J. Padberg and C.~H. Venner,
  \emph{Tribology letters}, 2016, \textbf{61}, 18\relax
\mciteBstWouldAddEndPuncttrue
\mciteSetBstMidEndSepPunct{\mcitedefaultmidpunct}
{\mcitedefaultendpunct}{\mcitedefaultseppunct}\relax
\EndOfBibitem
\bibitem[Roman \emph{et~al.}(2016)Roman, Valencia, and Franco]{Roman2016}
C.~Roman, C.~Valencia and J.~M. Franco, \emph{Tribology Letters}, 2016,
  \textbf{63}, 20\relax
\mciteBstWouldAddEndPuncttrue
\mciteSetBstMidEndSepPunct{\mcitedefaultmidpunct}
{\mcitedefaultendpunct}{\mcitedefaultseppunct}\relax
\EndOfBibitem
\bibitem[Walstra(1993)]{Walstra1993}
P.~Walstra, \emph{Cheese: Chemistry, physics and microbiology}, Springer, 1993,
  pp. 141--191\relax
\mciteBstWouldAddEndPuncttrue
\mciteSetBstMidEndSepPunct{\mcitedefaultmidpunct}
{\mcitedefaultendpunct}{\mcitedefaultseppunct}\relax
\EndOfBibitem
\bibitem[Liu \emph{et~al.}(2013)Liu, Zhang, Liu, Wang, and Jiang]{Liu2013}
H.~Liu, P.~Zhang, M.~Liu, S.~Wang and L.~Jiang, \emph{Advanced Materials},
  2013, \textbf{25}, 4477--4481\relax
\mciteBstWouldAddEndPuncttrue
\mciteSetBstMidEndSepPunct{\mcitedefaultmidpunct}
{\mcitedefaultendpunct}{\mcitedefaultseppunct}\relax
\EndOfBibitem
\bibitem[Yao \emph{et~al.}(2014)Yao, Ju, Yang, Wang, and Jiang]{Yao2014}
X.~Yao, J.~Ju, S.~Yang, J.~Wang and L.~Jiang, \emph{Advanced Materials}, 2014,
  \textbf{26}, 1895--1900\relax
\mciteBstWouldAddEndPuncttrue
\mciteSetBstMidEndSepPunct{\mcitedefaultmidpunct}
{\mcitedefaultendpunct}{\mcitedefaultseppunct}\relax
\EndOfBibitem
\bibitem[Urata \emph{et~al.}(2015)Urata, Dunderdale, England, and
  Hozumi]{Urata2015}
C.~Urata, G.~J. Dunderdale, M.~W. England and A.~Hozumi, \emph{Journal of
  Materials Chemistry A}, 2015, \textbf{3}, 12626--12630\relax
\mciteBstWouldAddEndPuncttrue
\mciteSetBstMidEndSepPunct{\mcitedefaultmidpunct}
{\mcitedefaultendpunct}{\mcitedefaultseppunct}\relax
\EndOfBibitem
\bibitem[Carretti \emph{et~al.}(2010)Carretti, Bonini, Dei, Berrie, Angelova,
  Baglioni, and Weiss]{Carretti2010}
E.~Carretti, M.~Bonini, L.~Dei, B.~H. Berrie, L.~V. Angelova, P.~Baglioni and
  R.~G. Weiss, \emph{Accounts of chemical research}, 2010, \textbf{43},
  751--760\relax
\mciteBstWouldAddEndPuncttrue
\mciteSetBstMidEndSepPunct{\mcitedefaultmidpunct}
{\mcitedefaultendpunct}{\mcitedefaultseppunct}\relax
\EndOfBibitem
\bibitem[Pianorsi \emph{et~al.}(2017)Pianorsi, Raudino, Bonelli, Chelazzi,
  Giorgi, Fratini, and Baglioni]{Pianorsi2017}
M.~D. Pianorsi, M.~Raudino, N.~Bonelli, D.~Chelazzi, R.~Giorgi, E.~Fratini and
  P.~Baglioni, \emph{Pure and Applied Chemistry}, 2017, \textbf{89},
  3--17\relax
\mciteBstWouldAddEndPuncttrue
\mciteSetBstMidEndSepPunct{\mcitedefaultmidpunct}
{\mcitedefaultendpunct}{\mcitedefaultseppunct}\relax
\EndOfBibitem
\bibitem[Bremond \emph{et~al.}(2011)Bremond, Cayer-Barrioz, and
  Mazuyer]{Bremond2011}
F.~Bremond, J.~Cayer-Barrioz and D.~Mazuyer, Congr{\`e}s fran{\c{c}}ais de
  m{\'e}canique, 2011\relax
\mciteBstWouldAddEndPuncttrue
\mciteSetBstMidEndSepPunct{\mcitedefaultmidpunct}
{\mcitedefaultendpunct}{\mcitedefaultseppunct}\relax
\EndOfBibitem
\bibitem[Noordover \emph{et~al.}(2016)Noordover, David, Fiddelaers, and Van
  Den~Kommer]{Noordover2016}
A.~Noordover, S.~David, F.~Fiddelaers and A.~Van Den~Kommer, \emph{Grease test
  kit and methods of testing grease}, 2016, US Patent 9,341,611\relax
\mciteBstWouldAddEndPuncttrue
\mciteSetBstMidEndSepPunct{\mcitedefaultmidpunct}
{\mcitedefaultendpunct}{\mcitedefaultseppunct}\relax
\EndOfBibitem
\bibitem[Rosenholm(2015)]{Rosenholm2015}
J.~B. Rosenholm, \emph{Advances in Colloid and Interface Science}, 2015,
  \textbf{220}, 8--53\relax
\mciteBstWouldAddEndPuncttrue
\mciteSetBstMidEndSepPunct{\mcitedefaultmidpunct}
{\mcitedefaultendpunct}{\mcitedefaultseppunct}\relax
\EndOfBibitem
\bibitem[Danino and Marmur(1994)]{Danino1994}
D.~Danino and A.~Marmur, \emph{Journal of colloid and interface science}, 1994,
  \textbf{166}, 245--250\relax
\mciteBstWouldAddEndPuncttrue
\mciteSetBstMidEndSepPunct{\mcitedefaultmidpunct}
{\mcitedefaultendpunct}{\mcitedefaultseppunct}\relax
\EndOfBibitem
\bibitem[Washburn(1921)]{Washburn1921}
E.~W. Washburn, \emph{Physical review}, 1921, \textbf{17}, 273\relax
\mciteBstWouldAddEndPuncttrue
\mciteSetBstMidEndSepPunct{\mcitedefaultmidpunct}
{\mcitedefaultendpunct}{\mcitedefaultseppunct}\relax
\EndOfBibitem
\bibitem[Mendez \emph{et~al.}(2009)Mendez, Fenton, Gallegos, Petsev, Sibbett,
  Stone, Zhang, and L{\'o}pez]{Mendez2009}
S.~Mendez, E.~M. Fenton, G.~R. Gallegos, D.~N. Petsev, S.~S. Sibbett, H.~A.
  Stone, Y.~Zhang and G.~P. L{\'o}pez, \emph{Langmuir}, 2009, \textbf{26},
  1380--1385\relax
\mciteBstWouldAddEndPuncttrue
\mciteSetBstMidEndSepPunct{\mcitedefaultmidpunct}
{\mcitedefaultendpunct}{\mcitedefaultseppunct}\relax
\EndOfBibitem
\bibitem[Gillespie(1958)]{Gillespie1958}
T.~Gillespie, \emph{Journal of Colloid Science}, 1958, \textbf{13},
  32--50\relax
\mciteBstWouldAddEndPuncttrue
\mciteSetBstMidEndSepPunct{\mcitedefaultmidpunct}
{\mcitedefaultendpunct}{\mcitedefaultseppunct}\relax
\EndOfBibitem
\bibitem[Hulst and van~de Hulst(1981)]{vdHulst1981}
H.~C. Hulst and H.~C. van~de Hulst, \emph{Light scattering by small particles},
  Courier Corporation, 1981\relax
\mciteBstWouldAddEndPuncttrue
\mciteSetBstMidEndSepPunct{\mcitedefaultmidpunct}
{\mcitedefaultendpunct}{\mcitedefaultseppunct}\relax
\EndOfBibitem
\bibitem[de~Meijer \emph{et~al.}(2000)de~Meijer, Haemers, Cobben, and
  Militz]{Meijer2000}
M.~de~Meijer, S.~Haemers, W.~Cobben and H.~Militz, \emph{Langmuir}, 2000,
  \textbf{16}, 9352--9359\relax
\mciteBstWouldAddEndPuncttrue
\mciteSetBstMidEndSepPunct{\mcitedefaultmidpunct}
{\mcitedefaultendpunct}{\mcitedefaultseppunct}\relax
\EndOfBibitem
\end{mcitethebibliography}
\bibliographystyle{rsc} 

\end{document}